\documentclass[twoside,twocolumn,9pt]{article}
\usepackage{extsizes}
\usepackage[super,sort&compress,comma]{natbib} 
\usepackage[version=3]{mhchem}
\usepackage[left=1.5cm, right=1.5cm, top=1.785cm, bottom=2.0cm]{geometry}
\usepackage{balance}
\usepackage{mathptmx}
\usepackage{sectsty}
\usepackage{graphicx} 
\usepackage{lastpage}
\usepackage[format=plain,justification=justified,singlelinecheck=false,font={stretch=1.125,small,sf},labelfont=bf,labelsep=space]{caption}
\usepackage{subcaption} % FOR MULTI-PANEL FIGURES
\usepackage{float}
\usepackage{fancyhdr}
\usepackage{fnpos}
\usepackage[english]{babel}
\addto{\captionsenglish}{%
  
}
\usepackage{array}
\usepackage{droidsans}
\usepackage{charter}
\usepackage[T1]{fontenc}
\usepackage[usenames,dvipsnames]{xcolor}
\usepackage{setspace}
\usepackage[compact]{titlesec}
\usepackage{hyperref}
\usepackage{epstopdf}%This line makes .eps figures into .pdf - please comment out if not required.

\definecolor{cream}{RGB}{222,217,201}

\begin{document}

\pagestyle{fancy}
\thispagestyle{plain}
\fancypagestyle{plain}{
%%%HEADER%%%
\renewcommand{\headrulewidth}{0pt}
}
%%%END OF HEADER%%%

%%%PAGE SETUP - Please do not change any commands within this section%%%
\makeFNbottom
\makeatletter
\renewcommand\LARGE{\@setfontsize\LARGE{15pt}{17}}
\renewcommand\Large{\@setfontsize\Large{12pt}{14}}
\renewcommand\large{\@setfontsize\large{10pt}{12}}
\renewcommand\footnotesize{\@setfontsize\footnotesize{7pt}{10}}
\makeatother

\renewcommand{\thefootnote}{\fnsymbol{footnote}}
\renewcommand\footnoterule{\vspace*{1pt}% 
\color{cream}\hrule width 3.5in height 0.4pt \color{black}\vspace*{5pt}} 
\setcounter{secnumdepth}{5}

\makeatletter 
\renewcommand\@biblabel[1]{#1}            
\renewcommand\@makefntext[1]% 
{\noindent\makebox[0pt][r]{\@thefnmark\,}#1}
\makeatother 
\renewcommand{\figurename}{\small{Fig.}~}
\sectionfont{\sffamily\Large}
\subsectionfont{\normalsize}
\subsubsectionfont{\bf}
\setstretch{1.125} %In particular, please do not alter this line.
\setlength{\skip\footins}{0.8cm}
\setlength{\footnotesep}{0.25cm}
\setlength{\jot}{10pt}
\titlespacing*{\section}{0pt}{4pt}{4pt}
\titlespacing*{\subsection}{0pt}{15pt}{1pt}
%%%END OF PAGE SETUP%%%

%%%FOOTER%%%
\fancyfoot{}
\fancyfoot[LO,RE]{\vspace{-7.1pt}\includegraphics[height=9pt]{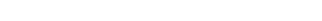}}
\fancyfoot[CO]{\vspace{-7.1pt}\hspace{13.2cm}\includegraphics{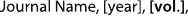}}
\fancyfoot[CE]{\vspace{-7.2pt}\hspace{-14.2cm}\includegraphics{head_foot/RF}}
\fancyfoot[RO]{\footnotesize{\sffamily{1--\pageref{LastPage} ~\textbar  \hspace{2pt}\thepage}}}
\fancyfoot[LE]{\footnotesize{\sffamily{\thepage~\textbar\hspace{3.45cm} 1--\pageref{LastPage}}}}
\fancyhead{}
\renewcommand{\headrulewidth}{0pt} 
\renewcommand{\footrulewidth}{0pt}
\setlength{\arrayrulewidth}{1pt}
\setlength{\columnsep}{6.5mm}
\setlength\bibsep{1pt}
%%%END OF FOOTER%%%

%%%FIGURE SETUP - please do not change any commands within this section%%%
\makeatletter 
\newlength{\figrulesep} 
\setlength{\figrulesep}{0.5\textfloatsep} 

\newcommand{\topfigrule}{\vspace*{-1pt}% 
\noindent{\color{cream}\rule[-\figrulesep]{\columnwidth}{1.5pt}} }

\newcommand{\botfigrule}{\vspace*{-2pt}% 
\noindent{\color{cream}\rule[\figrulesep]{\columnwidth}{1.5pt}} }

\newcommand{\dblfigrule}{\vspace*{-1pt}% 
\noindent{\color{cream}\rule[-\figrulesep]{\textwidth}{1.5pt}} }

\makeatother
%%%END OF FIGURE SETUP%%%

%%%TITLE, AUTHORS AND ABSTRACT%%%
\twocolumn[
  \begin{@twocolumnfalse}
{
%\includegraphics[height=30pt]%{head_foot/SM}
%\hfill\raisebox{0pt}[0pt][0pt]%{\includegraphics[height=55pt]{head_foot/RSC_LOGO_CMYK}}\\[1ex]
%\includegraphics[width=18.5cm]{head_foot/header_bar}
}\par
\vspace{1em}
\sffamily
\begin{tabular}{m{4.5cm} p{13.5cm} }

\; \; & \noindent\LARGE{\textbf{Re-entrant percolation in active Brownian hard disks$^\dag$}} \\%Article title goes here instead of the text "This is the title"
\vspace{0.3cm} & \vspace{0.3cm} \\

& \noindent\large{David Evans$^{a}$, José Martin-Roca$^{b,c}$, Nathan J. Harmer$^{a,d}$, Chantal Valeriani$^{b,c}$ and Mark A. Miller$^{a}$} \\%Author names go here instead of "Full name", etc.

\; \; & \noindent\normalsize{%
Non-equilibrium clustering and percolation are investigated in an archetypal model of two-dimensional active matter using dynamic simulations of self-propelled Brownian repulsive particles.  We concentrate on the single-phase region up to moderate levels of activity, before motility-induced phase separation (MIPS) sets in.  Weak activity promotes cluster formation and lowers the percolation threshold.  However, driving the system further out of equilibrium partly reverses this effect, resulting in a minimum in the critical density for the formation of system-spanning clusters and introducing re-entrant percolation as a function of activity in the pre-MIPS regime.  This non-monotonic behaviour arises from competition between activity-induced effective attraction (which eventually leads to MIPS) and activity-driven cluster breakup.  Using an adapted iterative Boltzmann inversion method, we derive effective potentials to map weakly active cases onto a passive (equilibrium) model with conservative attraction, which can be characterised by Monte Carlo simulations.  While the active and passive systems have practically identical radial distribution functions, we find decisive differences in higher-order structural correlations, to which the percolation threshold is highly sensitive.  For sufficiently strong activity, no passive pairwise potential can reproduce the radial distribution function of the active system.}
% Referee 2, comment 1

\end{tabular}

\end{@twocolumnfalse} \vspace{0.6cm} ]
%%%END OF TITLE, AUTHORS AND ABSTRACT%%

%%%FONT SETUP - please do not change any commands within this section
\renewcommand*\rmdefault{bch}\normalfont\upshape
\rmfamily
\section*{}
\vspace{-1cm}

%%%FOOTNOTES%%%

\footnotetext{\textit{$^{a}$~Department of Chemistry, Durham University, South Road, Durham DH1 3LE, United Kingdom}}
\footnotetext{\textit{$^{b}$~Departamento de Estructura de la Materia, Física Térmica y Electrónica, Universidad Complutense de Madrid, 28040 Madrid, Spain}}
\footnotetext{\textit{$^{c}$Grupo Interdisciplinar Sistemas Complejos, Madrid, Spain}}
\footnotetext{\textit{$^{d}$Current address: Department of Chemical Engineering and Biotechnology, West Cambridge Site, Philippa Fawcett Drive, Cambridge CB3 0AS, United Kingdom}}

\footnotetext{\dag~Electronic Supplementary Information (ESI) available: [details of any supplementary information available should be included here]. See DOI: 10.1039/cXsm00000x/}

\section{Introduction}

Active Brownian particles (ABPs) are one of the canonical models for describing and studying the properties of active matter.  In their basic form, ABPs are spheres or disks that are self-propelled by a force of constant magnitude.  Energy is dissipated by friction, and the motion (both translational and rotational) is subject to stochastic kicks.  Like all active matter, systems of ABPs are intrinsically out of equilibrium.  They are typically studied under steady-state conditions defined by the level of particle motility.

Despite the simplicity of the ABP model, it exhibits rich phase behaviour. In particular, purely repulsive ABPs can aggregate under certain conditions, leading to the coexistence of low- and high-density regions despite the absence of an explicit attractive force.\cite{Stenhammar,digregorio2018,JOSE} This phenomenon, known as motility-induced phase separation (MIPS), occurs at sufficiently high mean density and activity in two\cite{fily2012,redner2013a,Stenhammar,martin2022dynamical} and three dimensions.\cite{Stenhammar,wysocki2014}  It is a consequence of the symmetry-breaking induced by the activity and the collisions between particles. However, even before MIPS takes place, a significant degree of clustering occurs,\cite{ginot2018aggregation} to the extent that macroscopic networks of particles can arise without phase separation, as already observed in equilibrium systems with conservative attractive forces.\cite{poon1997,miller2003} As far as we are aware, the formation of these pre-MIPS structures in purely repulsive ABPs has not been studied in detail yet.

The mean cluster size in a given system depends on conditions such as temperature and density.  At the transition from a distribution of discrete clusters to a system-spanning network, the mean cluster size diverges, defining the percolation threshold.  Percolation theory has been used to study a range of problems where the onset of such connectivity results in a sudden change in a physical property, from the spread of forest fires\cite{vonniessen1986,vandenberg2021,perestrelo2022} to the enhancement of electrical conductivity in insulating materials.\cite{marsden2018,sandler2003,bauhofer2009}
In the latter case, the addition of conducting ``filler'' nanoparticles to an insulating matrix causes a large and sudden increase in the conductivity of the composite material when the fillers form a network that connects opposite boundaries of a macroscopic sample.
In soft matter, percolation is one of the signatures of gelation, which can influence rheological properties and the process of phase separation.\cite{Coniglio2004,poon1997,helgeson2014}  The ubiquity of colloidal gels in the food and pharmaceutical industries therefore makes percolation a phenomenon of wide-ranging importance in soft matter.\cite{joshi2014}

The critical density at which system-spanning clusters first appear depends both on the shape of the particles\cite{balberg1983,balberg1984,bug1985continuum, atashpendar2020} and on the specific interactions between them.\cite{bug1985,safran1985,miller2003,nigro2012}  For example, increasing the aspect ratio of hard nanoparticles from spheres to long rods (such as carbon nanotubes) causes the packing fraction at the percolation threshold to decrease dramatically, but a simple inverse relationship between aspect ratio and the critical density is only reached at very high aspect ratios.\cite{schilling2015}

Percolation theory has mostly been applied to explain the physical behaviour of equilibrium systems, but there are cases where the formation of system-spanning networks is important out of equilibrium.  For example, the manufacture of composite materials often involves stirring for dispersal of the filler,\cite{Grossiord06a} and flow or spin processes\cite{Du05a} for moulding.  Any non-equilibrium structure produced by these processes is ``frozen'' into the network if the material is then solidified.  Shearing a system containing carbon nanotubes causes them to align with the direction of the flow, impairing connectivity and raising the percolation threshold.\cite{finner2018,pihlajamaa2021}  However, any local aggregation caused by the shear can have the opposite effect.\cite{kwon2012}
When considering  spherical particles, where alignment is not a consideration, the percolation threshold can either increase or decrease depending on the ratio of hard-core to connectivity distance and the strength of the shear flow.\cite{pihlajamaa2022}

In the case of active matter, Levis and Berthier predicted a steady enhancement of percolation (i.e., a lowering of the critical density) with increasing activity in a kinetic Monte Carlo model of ABPs, albeit with a definition of pairwise connectivity that increased in range along with the level of motility.\cite{Levis14a}
More recently, Sanoria and coworkers examined percolation in a suspension of soft ABPs, demonstrating that the softness of the interactions between particles has a major influence on structure and phase behaviour.\cite{sanoria2022} For disks where the repulsive force is only linear in the pairwise separation, these authors concluded that increasing activity is equivalent to increasing softness.  At high activity, the dense MIPS state gives way to a percolating porous network, which is then destroyed by a further increase in activity.
Most recently of all, Caporusso and coworkers have studied the phase diagram of active dumbbells in three dimensions with strong, short-ranged attraction.\cite{Caporusso24a}  Rich phase behaviour arises in this system due to the additional ingredient of particle shape on top of activity and the dimensionality of the space.  Percolation occurs both in an arrested gel phase in the passive limit (zero activity) due to the sticky conservative attraction, and at sufficiently high density and activity, where the motility of the particles breaks up large clusters.
% Referee 1, point 1

Considering pusher-type microswimmers---a different class of active particles---a percolation transition has been detected with increasing density at a specific level of activity, not only in suspensions of squirmers but also of asymmetric dumbbells.\cite{schwarzendahl2022}  In the case of run-and-tumble particles, simulations on a two-dimensional lattice have shown re-entrant percolation as a function of translational diffusion rate at low reorientation rates.\cite{saha2024}
A system of passive colloidal particles in a liquid of active chiral \textit{E.~coli} exhibits the characteristics of a percolation phase transition, albeit with critical exponents differing from those expected of the equilibrium case. \cite{kushwaha2024}
Colonies of gliding \textit{M.~xanthus} bacteria exhibit collective motion once they have self-assembled into large clusters---effectively a percolation transition arising from the bacteria's motility.\cite{Peruani12a}  Percolation is also important in the context of electrical signalling networks in \textit{B.~subtilis}, where the percolation threshold for signal transmission across a community coincides with the optimal balance between the cost of firing to individual cells and the benefit to the bacterial community.\cite{Larkin18a}
%Referee 2, comment 2

The aims of the present article are two-fold.  Firstly, we study the mechanism of percolation before MIPS appears in a two dimensional suspension of repulsive ABPs, unravelling how the percolation threshold changes as activity is increased. 
Secondly, we explore whether the effect of activity on percolation can be modelled as a small perturbation of a system of passive particles.  Our route to an effective passive model is an adaptation of the iterative Boltzmann inversion technique,\cite{mcgreevy1988,reith2003} which is usually used to devise potentials for coarse-grained models in equilibrium, starting from more detailed atomistic simulations.

\section{Simulations Details}
\subsection{Brownian dynamics of active particles}

We simulate a system of $N$ disks of diameter $\sigma$ in a two-dimensional square box with edge $L$ and periodic boundary conditions in both directions.  The packing fraction cannot be precisely defined because the effective particle diameter depends on activity, while the conventional Barker--Henderson expression\cite{barker} for effective diameter relies on equilibrium conditions.  We therefore use the number density $\rho=N/L^2$ as a parameter to change the state of the system.\cite{rogel,JOSE}
 
Active particles move via Brownian dynamics with a self-propelling force on each particle that has its own rotational dynamics. The translational and rotational equations of motion for this model are
\begin{align}
\label{eq:motion}
& \dot{\vec{r}}_i = \frac{D_{\rm t}}{k_{\rm B} T} \left( \vec{F}_i +  F_{\rm a} \, \vec{n}_i \right) + \sqrt{2D_{\rm t}} \, \vec{\xi}_i, \\
& \dot{\theta}_i = \sqrt{2D_{\rm r}}\, \xi_{i,\theta},
\end{align}
where $\vec{F}_i=- \sum_{j\neq i} \nabla V(r_{ij})$ is the conservative force acting on particle $i$ and $V(r_{ij})$ is the inter-particle pair potential. $k_{\rm B}$ is Boltzmann's constant and $T$ the absolute temperature. $F_{\rm a}$ is the (constant) modulus of the self-propulsion force acting along the  orientation vector $\vec{n}_i$, which forms an angle $\theta_i$ with the positive $x$-axis, and $D_{\rm t}$ and $D_{\rm r}$ are the translational and rotational diffusion coefficients, respectively. The components of the random forces $\vec{\xi}_i$ and $\xi_{i,\theta}$ are white noise with zero mean and delta-correlation: $\langle\xi^{\alpha}(t)\xi^{\beta}(t')\rangle = \delta_{\alpha\beta} \delta(t-t')$, where $\alpha,\beta$ each represent $x$ or $y$ components, and $\langle\xi_{\theta}(t)\xi_{\theta}(t')\rangle = \delta(t-t')$. 
For equilibrium systems, the two diffusion coefficients follow a Stokes--Einstein relation for spherical particles (with diameter $\sigma$) \cite{StokesEinstein}: $D_{\rm r} = 3D_{\rm t}/\sigma^2$. However, that coupling no longer holds out of equilibrium.\cite{Navarro2015,berg1993random, ginot2018sedimentation,patteson2015running, aragones2018diffusion} In our work we  will treat $D_{\rm t}$ and $D_{\rm r}$ as independent control variables in line with previous numerical studies.\cite{fily2012athermal, JOSE,martin2022dynamical}

\noindent
For the repulsive core we use the pseudo-hard sphere\cite{PHS,baez2018using} (PHS) model, which has been shown to reproduce equilibrium\cite{espinosa} and out-of-equilibrium\cite{rosales,montero} characteristics of hard-sphere suspensions, while being continuously differentiable. The PHS pairwise potential has the form
\begin{equation}
    V_{\rm PHS}(r)  = \left\{ \begin{array}{ll}
        50 \left( \frac{50}{49} \right)^{49} \epsilon \left[ \left( \dfrac{\sigma}{r} \right)^{50} -  \left(\dfrac{\sigma}{r} \right)^{49} \right] + \epsilon, & r< \left(\frac{50}{49}  \right)  \sigma \\
        0, & r \geq \left(\frac{50}{49}  \right)  \sigma
    \end{array}  \right.
     \label{eq:PHS}
\end{equation}
where $r$ is the centre-to-centre distance, $\sigma$ is the particle diameter and $\epsilon$ sets the energy scale.  We will express all lengths and energies in terms of $\sigma$ and $\epsilon$, respectively.  Introducing $m$ as the mass of a disk, we may also define a natural unit of time $\tau_{\rm PHS}=\sqrt{m\sigma^2/\epsilon}$.

Our goal is to study clustering and percolation before MIPS appears.  To generate configurations  for the percolation analysis, we simulate the system with 
%a modified version of 
LAMMPS\cite{LAMMPS} (Large-scale Atomic/Molecular Massively Parallel Simulator). We initiate the system with particles located on a square lattice and run for $10^6$ steps with a time step of $5\times10^{-6} \tau_{\rm PHS}$ to ensure the system reaches steady state.  The simulation is run for a further $2\times10^8$ steps, saving the particles' coordinates frequently, but with sufficient time between frames to ensure that the sampled configurations are independent.

\noindent
As a measure of the degree of activity, we use the P\'eclet number $\mathrm{Pe}$, i.e., the dimensionless ratio between advective and diffusive transport, defined as:
\begin{equation}
\mathrm{Pe} = \frac{3 v \, \tau_{\rm r}}{\sigma}= \frac{3 F_{\rm a} D_{\rm t}}{\sigma k_{\rm B} T D_{\rm r}},
\label{eq:Pe}
\end{equation}
where $v=F_{\rm a} D_{\rm t}/k_{\rm B}T$ is the self-propulsion velocity and $\tau_{\rm r} = 1/D_{\rm r}$ the reorientation time.
For the results presented in the main article, the activity is varied by changing $D_{\rm r}$ while fixing $F_{\rm a} = 24\epsilon/\sigma$, $k_{\rm B}T = 1.5\epsilon$ and $D_{\rm t} = (k_{\rm B} T /\epsilon)\sigma^2/\tau_{\rm PHS}$. The choice of these values is based on the work of Stenhammar\cite{Stenhammar} and other studies that deploy the potential in Eq.~(\ref{eq:PHS}).\cite{JOSE}  In ESI Section 1, we test the effect of controlling {\rm Pe} by varying $F_{\rm a}$ rather than $D_{\rm r}$.

\subsection{Monte Carlo of passive particles}

We will compare the structure of the active system to that of a passive (equilibrium) model with an effective interaction potential that is to be deduced in Section \ref{sec:ibi}.  The passive system is most easily simulated using standard canonical Monte Carlo methods.  Starting from an initial configuration of particles located on a square lattice, we perform stochastic trial displacements and test for acceptance of the move using the Metropolis algorithm,\cite{metropolis1953} with a target of 50\% of the moves being accepted.  We simulate for $5\times10^4$ MC sweeps to equilibrate and then gather statistics over a further $5\times10^5$ MC sweeps, saving the configuration every 1000 sweeps to produce 500 independent configurations.

\subsection{Detecting percolation}

The percolation threshold is defined as the critical set of conditions at which the mean cluster size diverges.  Beyond this point in the macroscopic limit, an ``infinite'', system-spanning cluster is always present.  In our simulations, a cluster is defined as a group of connected disks.  In keeping with general studies of continuum percolation, two disks are taken to be connected when their centres are separated by less than a threshold,\cite{bug1985} which we take to be $\lambda\sigma$, setting the range parameter $\lambda=1.3$.  This value of $\lambda$ allows a large range of densities and P{\'e}clet number to be studied without entering the MIPS region\cite{JOSE}.  The sensitivity of the results to $\lambda$ is tested in ESI Section 2.

To detect percolating clusters, we use a strict wrapping criterion,\cite{skvor2007,seaton1987} where a cluster is only counted as percolating if it connects to its own images through the periodic boundary conditions of the simulation cell, thereby generating an infinite structure.  This is a more stringent requirement than the cluster simply having a dimension large enough to connect opposite sides of the cell.  Wrapping clusters are detected by building a contiguous image of the cluster that contains exactly one image of each disk in the cluster, and then testing for pairs of particles that are not directly in contact in the isolated cluster, but become connected when the periodic boundary conditions are considered.  If the cluster contains one or more such pairs, then it wraps through the periodic boundaries, otherwise it is isolated and finite.\cite{miller2003}

The percolation probability, $R(L,\rho)$ is the probability that a randomly chosen configuration in a trajectory contains a percolating cluster in any direction in a simulation cell of linear dimension $L$.  This is calculated by counting the number of configurations that contain a wrapping cluster and dividing by the total number of configurations in the trajectory.

In the macroscopic limit, the percolation transition is characterised by a step change in $R(L,\rho)$ from 0 to 1 at a critical density $\rho=\rho_{\rm c}$.  However finite-size effects broaden this transition in a well-defined way for the equilibrium case.  One way to establish the value of $\rho_{\rm c}$ is to find the density at which the curves of $R(L,\rho)$ as a function of $\rho$ intersect for different values of $L$.  Alternatively, lines of constant $R$ can be extrapolated to infinite $L$, where they should converge at $\rho_{\rm c}$.  How this is done in practice is detailed in the next section.

\section{Results and Discussion}

\subsection{Percolation of active disks}

\begin{figure}
    \centering
    \includegraphics[width=\linewidth]{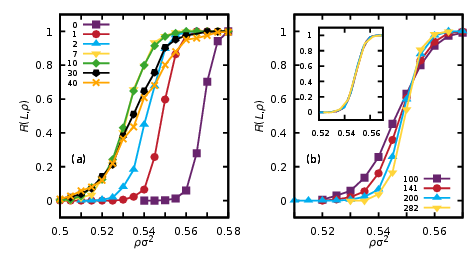}
    \caption{(a) Percolation probability $R(L,\rho)$ versus density $\rho$ for P{\'e}clet numbers in the range $0\le{\rm Pe}\le40$ for simulations of ABP in a box of linear size $L=200\sigma$.
    (b) Percolation probability as a function of density for weak activity, ${\rm Pe} = 1$ at four different linear box sizes.  The four curves intersect at the same density, corresponding to the percolation threshold.  The inset shows the results of scaling the curves onto the $L=200$ curve using the known scaling exponent for passive systems in two dimensions $\nu=4/3$.}
    \label{fig:probabilities}
\end{figure}

For a given value of the P{\'e}clet number and linear box dimension $L$, we locate the range of density over which the percolation probability $R(L,\rho)$ rises from $0$ to $1$.  The resulting curves for $0\le{\rm Pe}\le40$ at $L=200\sigma$ are shown in Fig.~\ref{fig:probabilities}a.  We immediately see that the introduction of a small amount of activity causes a shift in the percolation response to lower density, meaning that activity is promoting percolation.
Increasing the activity further, the trend continues up to approximately ${\rm Pe}=7$, where the shift stops, and increasing the activity further causes the transition to widen.

For each P{\'e}clet number, we then perform simulations at four box sizes, $L/\sigma=100,141,200,282$.
Figure \ref{fig:probabilities}b shows how $R(L,\rho)$ sharpens with increasing $L$ in the case of ${\rm Pe}=1$ and how this can be used to identify the percolation threshold as the common point at which all these curves intersect.
In passive systems,\cite{stauffer1994} the width of the response $R(L,\rho)$ is proportional to $L^{-1/\nu}$ where the value of the critical exponent $\nu$ in two dimensions is $4/3$, and curves for different $L$ should collapse onto each other if plotted as a function of the scaled density $x=(\rho-\rho_{\rm c})L^{1/\nu}$.  The inset of Fig.~\ref{fig:probabilities}b shows that these equilibrium scaling and collapse properties still apply in our weakly active, non-equilibrium system of ABPs.  Preservation of the equilibrium values of critical exponents has also been observed in very soft active particles in a different regime of activity in recent work by Sanoria et al.\cite{sanoria2022}

\begin{figure}
    \centering
    \includegraphics[width=\linewidth]{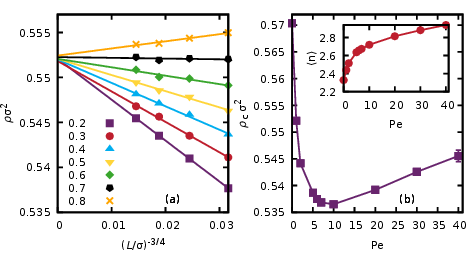}
    \caption{(a) The value of density $\rho$ for a given percolation probability $p=0.2, 0.3\dots 0.8$ versus $L^{-1/\nu}$ for activity ${\rm Pe}=1$ and $\nu=4/3$.  The lines are best fits through points with the same value of $p$.  The intercept of the lines with the vertical axis gives $\rho_{\rm c}$ for infinite box size and the spread of the intercepts gives the uncertainty in the threshold. (b) The percolation threshold as a function of P{\'e}clet number. Most error bars are smaller than the symbols. The inset of shows the monotonically increasing average coordination number with ${\rm Pe}$ at the percolation threshold.}
    \label{fig:thresholds}
\end{figure}

To calculate the percolation threshold, we identify the density at which each percolation probability curve equals a series of probabilities, $R(L,\rho)=p$ for $0.2\le p\le 0.8$ in steps of $\Delta p=0.1$, by linear interpolation between measured points on the curves.  All these densities should converge on the percolation threshold $\rho_{\rm c}$ as $L$ increases, since $R$ tends to a step function, $\lim_{L\to\infty}R(L,\rho)=\Theta(\rho-\rho_{\rm c})$.  Fig.~\ref{fig:thresholds}a shows lines of density at constant $R(L,\rho)=p$ as a function of $L^{-1/\nu}$ with $\nu=4/3$.  Extrapolation of the lines for different $p$ to $L^{-1/\nu}=0$ gives the percolation threshold $\rho_{\rm c}$ in the macroscopic limit.  We take the spread of the intercept values as a measure of the statistical uncertainty in $\rho_{\rm c}$.

We use this method to calculate the percolation threshold as a function of P{\'e}clet number, as reported in Figure \ref{fig:thresholds}b. As shown in the figure, a small amount of activity rapidly reduces the percolation threshold until a minimum is reached around ${\rm Pe} = 10$, after which increasing the activity further results in an increase in the percolation threshold.  This nonmonotonic behaviour creates re-entrant percolation as a function of increasing activity for $0.5365 < \rho\sigma^2 < 0.5455 $.  Analogous results for ABPs where activity is controlled by varying the propulsion force $F_{\rm a}$ rather than the rotational diffusion constant $D_{\rm r}$ are presented in the Electronic Supplementary Information (ESI).

We stop the percolation analysis at ${\rm Pe}=40$, which is just below the MIPS boundary.\cite{JOSE}  After phase separation, droplets of the dense phase are always fully connected, so the nature of percolation changes completely to a question of whether the droplet itself is large enough to span the simulation cell.

The mean coordination number $\langle n\rangle$ over all disks is far from being an invariant at the percolation transition, as shown in the inset of Figure \ref{fig:thresholds}b, rising by almost 30\% over the range of activities studied.  The rise is steepest when weak activity is introduced to the reference equilibrium model.  Tracking $\langle n\rangle$ along a line of constant density $\rho=0.54\sigma^{-2}$ rather than along the percolation threshold, we still see a monotonic rise, even as the system leaves the re-entrant region and stops percolating at higher ${\rm Pe}$ (Fig.~S6 of the ESI).  Hence, the increasing mean coordination of individual particles reflects greater internal networking within clusters rather than connectivity of the system on a larger scale.  

\begin{figure}[]
    \centering
    \captionsetup[subfigure]{justification=centering}
    \begin{subfigure} [t]{0.32\linewidth}
        \centering
        \includegraphics[width=\linewidth]{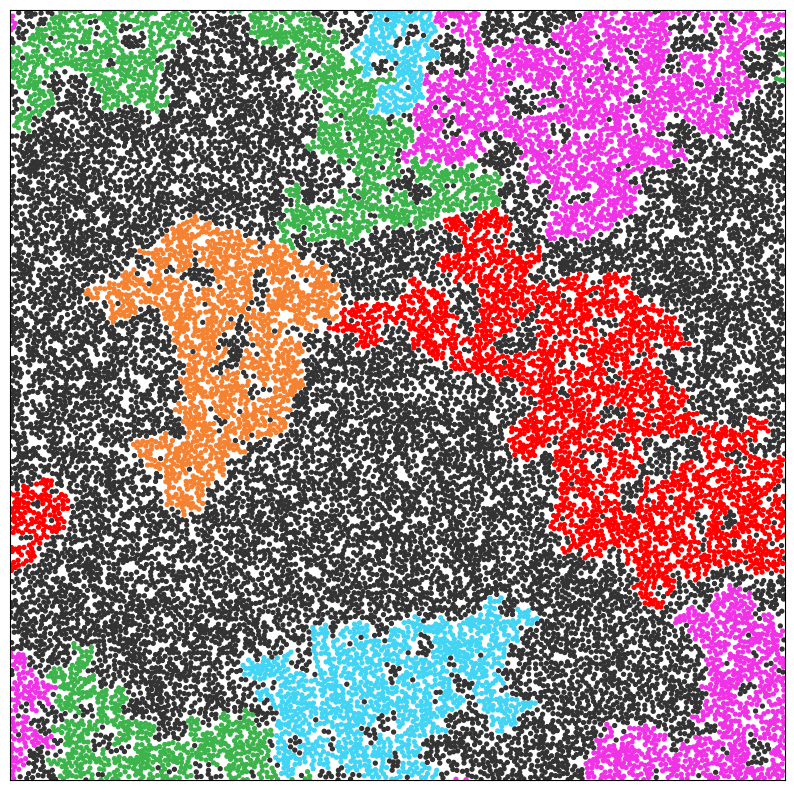}
        \caption{}
        \label{fgr:snapshotPe1}
    \end{subfigure}
    \begin{subfigure} [t]{0.32\linewidth}
        \centering
        \includegraphics[width=\linewidth]{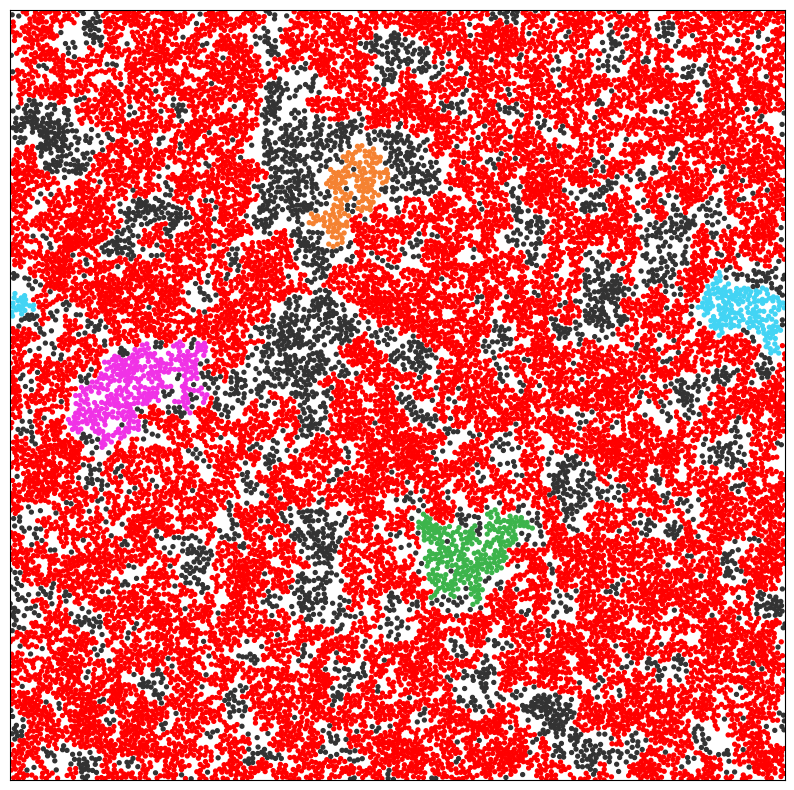}
        \caption{}
        \label{fgr:snapshotPe10}
    \end{subfigure}
    \begin{subfigure} [t]{0.32\linewidth}
        \centering
        \includegraphics[width=\linewidth]{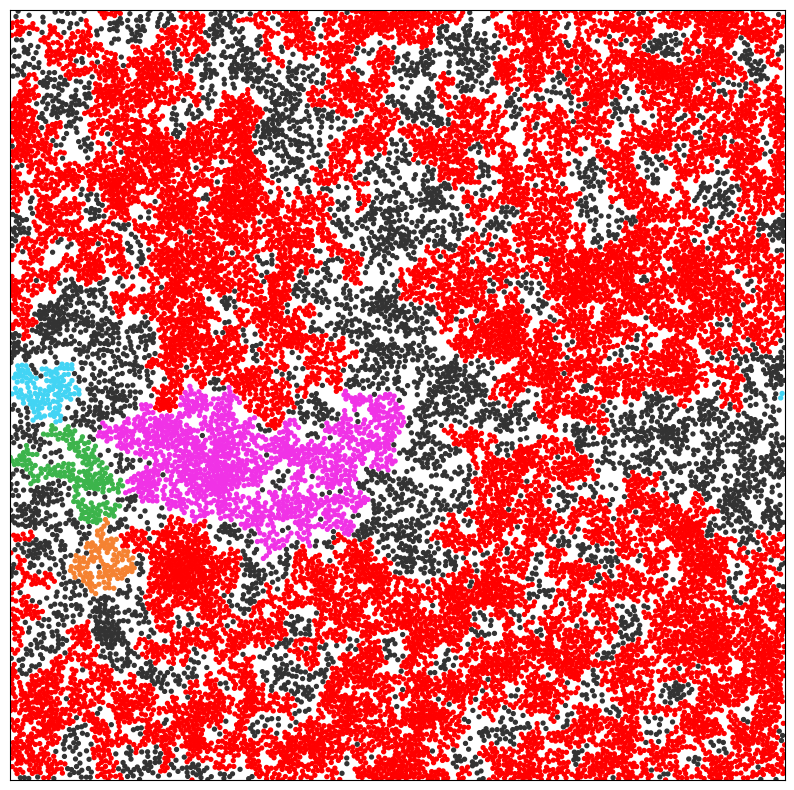}
        \caption{}
        \label{fgr:snapshotPe40}
    \end{subfigure}
    \caption{Snapshots of the $L=200\sigma$ system at a density of $\rho=0.540\sigma^{-2}$ and (a) ${\rm Pe} = 1$, (b) ${\rm Pe} = 10$, (c) ${\rm Pe} = 40$. The five largest clusters are coloured red (largest of all), magenta, green, cyan and orange.  Black particles belong to clusters that are not in the five largest.  Only panel (b) contains a percolating cluster.}
    \label{fig:snapshots}
\end{figure}

The evolution towards intra-cluster connections implies a growing local inhomogeneity of the fluid structure.  This effect is visible as an increasingly patchy texture with activity (but without phase separation) in the snapshots in Fig.~\ref{fig:snapshots}.  The local density at a given particle can be quantified by the inverse area of its cell in a Voronoi tesselation of the system.  The increasing patchiness registers as a broadening of the distribution of this local density with P{\'e}clet number (see ESI Fig.~S9).

As in any case of continuum percolation, decreasing the connectivity range $\lambda$ shifts the percolation threshold to higher densities, since fewer pairs qualify as neighbours.  As shown in Fig.~S2 of the ESI, we observe pre-MIPS re-entrant percolation down to $\lambda=1.2$.  Below this range, the minimum in the percolation threshold is driven towards the density of the MIPS transition, which then intervenes.  Conversely, re-entrance persists on increasing the connectivity range at least as far as $\lambda=1.4$, with the minimum occurring at decreasing Pe as the range lengthens.
% Referee 2, comment 3

\subsection{Effective interaction potentials\label{sec:ibi}}
To understand the re-entrant behaviour of the percolation threshold in Fig.~\ref{fig:thresholds}b, we test whether these weakly active systems can be modelled as a perturbation of a passive, equilibrium system with an effective potential to account for the structural effects of activity.  Schemes for extracting effective potentials specifically for active matter have been devised by other authors, including approximate first-principles approaches by Farage et al.\cite{farage2015} (further investigated by Rein and Speck\cite{Rein2016}) and by Cameron et al.\cite{Cameron23a}, and machine-learning approaches by Bag and Mandal\cite{bag2021} and by Ruiz-Garcia.\cite{ruizgarcia2024}

In the context of ABP percolation, we want to capture the structural correlations of the pre-MIPS steady state in as much detail as possible rather than deriving an equation of state or predicting the MIPS transition.  With these priorities in mind, we use a modified iterative Boltzmann inversion (IBI) method\cite{mcgreevy1988,reith2003} to find effective pairwise potentials from the radial distribution functions (RDFs) measured in the active simulations.  IBI is usually used to deduce coarse-grained potentials for equilibrium simulations, starting from more detailed atomistic simulations.  It relies on Henderson's theorem,\cite{henderson1974} which shows that if a RDF is the result of pairwise interactions between particles then the underlying pair potential $V(r)$ is unique.  IBI therefore compares the RDF of the true system being modelled with the RDF returned from a simulation using a trial potential, expressed in piecewise linear form.  The difference between the two RDFs is used to modify the trial potential.  Further iterations of comparison and modification should lead to convergence of the two RDFs and (according to Henderson) the correct pair potential.  To the best of our knowledge, IBI has not previously been used to extract effective pairwise potentials for active matter.

We already know the conservative PHS contribution of Eq.~(\ref{eq:PHS}) to the disk interactions, so we split our putative effective potential $V_{\rm eff}(r)$ into a sum of the PHS potential and the contribution $V_{\rm active}(r)$ from the activity:
\begin{equation}
    V_{\textrm{eff}}(r) = V_{\textrm{PHS}}(r) + V_{\textrm{active}}(r),
\end{equation}
The unknown $V_{\textrm{active}}$ is the only part of the potential that is changed via the IBI procedure as follows:
\begin{equation}
    V_{\textrm{active}}(r) \leftarrow V_{\textrm{active}}(r) + k_{\textrm{B}}T \ln \left[\frac{g(r)}{g_{\textrm{target}}(r)}\right],
\end{equation}
where $g(r)$ and $g_{\textrm{target}}(r)$ are the radial distribution functions obtained from a Monte Carlo simulation using the current trial potential $V_{\rm eff}(r)$ and the dynamic simulations of the active system respectively.
$V_{\rm active}(r)$ is tabulated on a grid with spacing $\delta r=0.01\sigma$, matching the bin width of $g(r)$.  Linear interpolation is used between the tabulated points, while $V_{\rm PHS}(r)$ is evaluated directly from Eq.~(\ref{eq:PHS}) to obtain $V_{\rm eff}(r)$ for the Monte Carlo simulations.
We truncate the potentials at $r = 3\sigma$, which is large enough to capture the important structural information and to avoid a prominent discontinuity at the cutoff.
It has been shown that increasing the cutoff in IBI does not necessarily capture any additional information and can make convergence worse.\cite{reith2003}

To assist convergence, we initialise $V_{\rm active}(r)$ to 0 for the lowest activity level (${\rm Pe} = 1$) and then seed each successively higher level of activity with the converged result of the previous case.  This approach provides the best possible starting point for each optimisation and leads to better convergence than an initial approximate inversion of $g(r)$, as usually used in IBI in the context of coarse-grained potentials.  We terminate the iterative process when the total squared deviation between the target active and effective RDFs stops decreasing and reaches a decisive plateau (see ESI Fig.~S3).

\begin{figure}[h!]
    \centering
    \includegraphics[width=\linewidth]{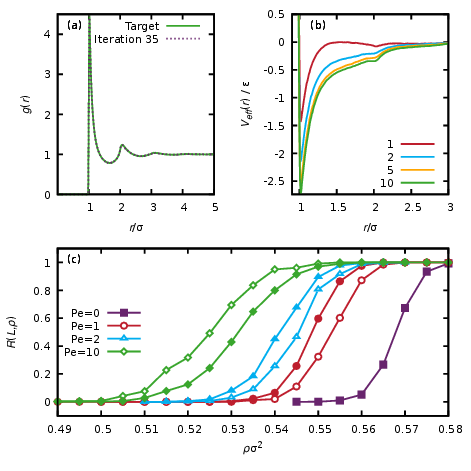}
    \caption{(a) The radial distribution function for active dynamics simulations at ${\rm Pe} = 2$, $\rho=0.54\sigma^{-2}$ and the result of passive Monte Carlo simulations with an effective potential after 35 iterations of IBI. (b) The potentials produced by IBI at different values of the P{\'e}clet number as labelled. (c) Percolation probability curves for active systems (filled symbols) and passive systems with effective potentials from IBI (open symbols), system size $L=200\sigma$. Each colour represents a different activity level. The Monte Carlo result for passive PHS ({\em i.e.}, ${\rm Pe} = 0$) is included for reference.}
    \label{fig:ibi}
\end{figure}

Figure \ref{fig:ibi}a shows a comparison of $g(r)$ for ${\rm Pe} = 2$ at $\rho = 0.54\sigma^{-2}$, and the result of the modified IBI procedure.  After 35 iterations, the two RDFs are indistinguishable within the residual statistical noise.  Hence, IBI is indeed able to return a pair potential that accurately reproduces the RDF of the active fluid.
Figure \ref{fig:ibi}b shows the resulting potentials $V_{\rm eff}(r)$ from the modified IBI procedure for four different P{\'e}clet numbers at $\rho = 0.54\sigma^{-2}$.  The effective potentials share common features, with a sharp attractive well at contact, which deepens with increasing activity as expected.

There is also a shallower secondary feature at $r=2\sigma$ before a tail to 0, which is related to the second peak in $g(r)$ at that radius.  This feature already distinguishes the ABP effective potential from the potential of a simple passive fluid such as Lennard-Jones (LJ), where $g(r)$ may also show several peaks but the potential only has a single minimum.  The second and later peaks in the LJ (and similar) radial distribution functions are the result of correlations between successive shells of neighbours, and no additional features are required in the potential energy function at the positions of the peaks in $g(r)$.  IBI correctly returns a smoothly decaying LJ potential with increasing $r$ when presented with a multi-peak $g(r)$ to invert under passive conditions.\cite{reith2003}  In contrast, the fact that the ABP radial distribution function generates a second feature in the effective potential implies that indirect correlations of next-nearest neighbours interacting through the primary minimum of $V_{\rm eff}(r)$ would not be sufficient to reproduce the detailed structure of the active fluid, even at the weakest levels of activity studied here.
% Referee 2, point 4

Beyond ${\rm Pe} = 10$, IBI fails to return a $V_{\rm eff}(r)$ that accurately reproduces $g_{\rm target}(r)$.  Henderson's theorem guarantees that a given RDF originates from a unique pair potential if only pairwise interactions are at play.\cite{henderson1974}  However, it does not guarantee that any RDF can be produced by a pairwise potential.  Our simulations of ABPs at moderate levels of activity provide examples of RDFs for which a generating pair potential does not seem to exist.  We may infer that at ${\rm Pe}=10$, our ABP system is already showing the effects of many-body and/or directional interactions in the RDF (which is still a pairwise correlation function).  These effects cannot be fully imitated in a system with conservative, isotropic, pairwise interactions.
Unfortunately, the apparent nonexistence of a faithful $V_{\rm eff}(r)$ as MIPS is approached means that we cannot gain insight into that transition through the evolution of the effective potential.
% Referee 1, point 2

Using the effective potentials for weak activity, where the RDF is accurately reproduced, we perform larger Monte Carlo simulations and calculate the percolation probabilities, which are shown in Fig.~\ref{fig:ibi}c (open symbols).  The simulations using the effective potentials produce a decisively different percolation response from their active counterparts, even for the weakest activity studied of ${\rm Pe}=1$.  For ${\rm Pe} = 1$ and $2$, percolation in the passive simulations is shifted to higher density, whereas for ${\rm Pe} = 10$ it is shifted to lower density.
We have also calculated other structural quantities, such as the average coordination number, radius of gyration and cluster size distribution, which show small deviations when comparing the passive and active systems.  The details of these calculations and the results are shown in the ESI.

The comparison between the active and passive cases immediately shows that percolation is extremely sensitive to details of a fluid's structure.  Each pair of passive and active simulations in Fig.~\ref{fig:ibi}c has practically identical RDFs, but distinct clustering.  It is already known that RDFs can be insensitive to the precise form of the potential,\cite{wang2020} despite the formal uniqueness theorem.\cite{henderson1974}  Here, however, we are comparing other structural aspects of fluids where the RDFs are already faithfully reproduced.  The differences in percolation indicate that there must be structural differences in the two systems beyond the pairwise correlation function $g(r)$.

Following the work of Torquato et al. and Wang et al.,\cite{torquato1990,wang2020} we calculate the conditional nearest neighbour distribution, $G_V(r)$, which provides a signature of any many body effects in the system.  $2\pi r\rho G_V(r)\,dr$ is the probability that, given a circular region containing no particle centres, a particle centre lies in a shell of thickness $dr$ around this empty region.  By construction, $G_V(r)$ up to the radius of a hard particle (approximately $r=\sigma/2$ in our model) is determined solely by the particle density, since the density sets the probability that a point inside a disk (which is inaccessible to the centres of other particles) has been selected.  Beyond that point, $G_V(r)$ can in principle be expressed in terms of many-body correlation functions.\cite{torquato1990}  ESI Fig.~S10 shows that there is a distinct difference between $G_V(r)$ for the active and passive systems, despite the ``structural degeneracy'' of the systems at the pairwise level.\cite{Stillinger19a}  The deviation of the $G_V(r)$ curves implies that many-body correlations exist in the structure of ABPs from very low P{\'e}clet number and therefore that the structural effects of even weak activity cannot be captured in full by an effective pairwise potential.

Setting aside the detailed mapping of active disks onto an effective passive system, we may still ask whether re-entrant percolation can arise from simple pairwise attraction at equilibrium as a function of the strength of the attraction with respect to the thermal energy.  We have performed equilibrium MC simulations on square-well disks with range $\lambda_{\rm SW}\sigma$ and depth $\epsilon_{\rm SW}$,
\begin{equation}
V_{\rm SW}(r) =
\begin{cases}
    \infty & r<\sigma \\
    -\epsilon_{\rm SW} & \sigma\le r<\lambda_{\rm SW}\sigma \\
    0 & r\ge\lambda_{\rm SW}\sigma,
\end{cases}
\end{equation}
to test how clustering varies with temperature.  Setting both the range of the attraction $\lambda_{\rm SW}$ and the connectivity criterion $\lambda$ equal to $1.3$ (matching our ABP simulations), we find a monotonically decreasing percolation threshold with decreasing temperature until gas--liquid phase separation intervenes (see ESI Fig.~S4).  This qualitative behaviour persists for different values of $\lambda=\lambda_{\rm SW}$.  Nevertheless, for an estimate of the square-well ranges corresponding to the activity-induced potentials in Fig.~\ref{fig:ibi}b, we have evaluated the second virial coefficient $B_2$ and determined the value of $\lambda_{\rm SW}$ that would return the same value if the well depths are also matched.  Beyond about ${\rm Pe}=2$, the effective square-well range settles down at $\lambda_{\rm SW}=1.23$ (ESI Section S5).

Data in an early study of percolation of square-well disks\cite{heyes1989} imply that re-entrance might arise if the range of attraction is decoupled from the connectivity distance and set to a significantly shorter value, $\lambda_{\rm SW}<\lambda$, although the authors did not draw attention to this point.  We have investigated this possibility with the benefit of more powerful computers and the full treatment of finite-size scaling as deployed earlier in the present work, but have not found any combination of $\lambda$ and $\lambda_{\rm SW}$ that gives rise to re-entrance.  Figure S4 of the ESI shows some example results for $\lambda_{\rm SW}<\lambda$ at $\lambda=1.3$.  Hence, the re-entrance seen in ABPs does seem to be an inherent and non-trivial result of the activity.

\section{Conclusions}

Our simulations show that weak activity promotes percolation in systems of disks where the conservative interactions are purely repulsive.  This initial response is readily understood in terms of the effective attraction induced by the activity, which increases the connectivity of the particles, leading to divergence of the linear cluster size at lower number density than in the underlying passive system.  With our modified iterative Boltzmann inversion scheme, we have extracted the detailed form of an effective pair potential that reproduces the structure of the active fluid as far as two-body correlations, while showing that higher-order correlations are also present.

Less expected is the nonmonotonic effect of activity, which is seen as a minimum in the percolation threshold around ${\rm Pe}=10$ and suppression of percolation relative to that point upon further increasing the activity, despite the continually increasing mean coordination number.  This effect can be related to a study of the phase behaviour of ABPs with LJ attraction by Redner, Baskaran and Hagan (RBH)\cite{redner2013}.  In that work, strong clustering and even phase separation occur in the absence of activity at sufficiently low temperature due to the conservative LJ forces.  Introducing moderate activity breaks up the clusters by driving the particles out of their potential energy minima, reversing the effects of the LJ attraction and yielding a single active phase.  Stronger activity then leads to phase separation by a different mechanism, MIPS, just as it does even for hard ABPs.  RBH's paper concentrates on this re-entrance of the phase behaviour.  In our work, which is confined to the pre-MIPS regime, the ABPs lack conservative attraction.  Hence, unlike the system studied by RBH, there is no phase separation or attraction-enhanced clustering at zero activity.  Without conservative attraction to obscure it, we therefore initially see the effects of activity-induced attraction by increased clustering and percolation at small ${\rm Pe}$.  However, analogous to the observations on phase separation by RBH,\cite{redner2013} increased activity reverses the effects of attraction, breaking up the clusters and suppressing percolation.  The key difference for the hard ABPs in our work is that the initial attraction is itself induced by weak activity before the clustering effects that it induces are partially destroyed by stronger activity.  Like the LJ system studied by RBH, a MIPS transition arises for ABPs at even higher activity\cite{JOSE} but we have deliberately confined our study of percolation to the one-phase, pre-MIPS regime.

Effective pairwise potentials can be deduced for ABPs up to ${\rm Pe}=10$, enabling the radial distribution functions of the active system to be reproduced in passive MC simulations.  However, comparison of the active and mapped-passive simulations shows that higher-order correlations are present even for very weak activity.  These many-body effects are effectively invisible at the two-body level of description in the sense that the radial distribution functions are identical to within the statistical noise, which itself is minimal.  Our use of IBI to extract effective pairwise potentials from active simulations is just one example of the broader challenge of deducing interactions from trajectories,\cite{Stones19a,Stones23a} which may originate from simulation or experiment, in or out of equilibrium, and may involve directional or many-body contributions in addition to isotropic pairwise terms.  Work towards a more general and sophisticated treatment of these interactions is currently underway.

\section*{Conflicts of interest}
There are no conflicts of interest to declare.

\section*{Acknowledgements}
Some of the work has been performed under the Project HPC-EUROPA3 (INFRAIA-2016-1-730897), with the support of the EC Research Innovation Action under the Horizon 2020 Programme.
DE acknowledges funding from the EPSRC SOFI$^2$ Centre for Doctoral Training (Grant EP/S023631/1). CV acknowledges funding IHRC22/00002 and PID2022-140407NB-C21 from MINECO.

\section*{Data Availability Statement}
Raw data for all figures in the main article and the ESI document are available in a zip file as part of the ESI.

\balance

\bibliography{rsc} 
\bibliographystyle{rsc} 

\end{document}